\documentclass[preprint,12pt,onecolumn]{elsarticle}%

\usepackage{amssymb}
\usepackage{amsfonts}
\usepackage{amsmath}
\usepackage{graphicx}
\usepackage{dcolumn}
\usepackage{bm}
\usepackage{lineno}
\usepackage{revsymb}%
\usepackage{hyperref}
\journal{Physica B}

\begin{document}
\begin{frontmatter}

\title{Ratchet universality and optimal suppression of shot noise in
biharmonically-driven tunnel junctions}

\author{Pedro J. Mart\'{\i}nez}
\address{Departamento de F\'{\i}sica Aplicada, E.I.N.A., Universidad
de Zaragoza\\
E-50018 Zaragoza, Spain and \\Instituto de Nanociencia y Materiales de
Arag\'{o}n (INMA), CSIC-Universidad de Zaragoza\\ E-50009 Zaragoza,
Spain\\ icmat1@unizar.es
}

\author{Ricardo Chac\'{o}n}
\address{Departamento de F\'{\i}sica Aplicada, E. I. I., Universidad de
Extremadura\\ Apartado Postal 382, E-06006 Badajoz, Spain and\\
Instituto de
Computaci\'{o}n Cient\'{\i}fica Avanzada (ICCAEx), Universidad de
Extremadura\\
E-06006 Badajoz, Spain.\\ rchacon@unex.es
}

\begin{abstract}
This work discusses two retrodictions of the ratchet
universality (RU) law
which explain previous numerical and experimental findings concerning the diode
effect in conventional superconducting tunnel-junctions (TJ) and
the controlled suppression of electron-hole pair generation in a TJ,
both
under biharmonic driving fields.
We demonstrate that the interplay between the RU law and the
Heisenberg's uncertainty principle leads to a
universal scenario of optimal quantum transport. Specifically, we
demonstrate that the RU driving field maximizes the diode's
efficiency while yielding a maximal rectification range for the
supercurrent, on the one hand, and optimally reduces the excess
quantum noise with
respect to the dc noise level, thus allowing for the efficient
production of
nonclassical photonic states. These results suggest that the RU law
seems essential for any \textit{optimal} application of the
ratchet effect, particularly in the contexts of superconducting
integrated power electronics, electron quantum optics, and quantum
computing.
\end{abstract}


\begin{keyword}
Ratchet Universality, superconducting tunnel-junctions, shot noise,
quantum computing.
\end{keyword}

\end{frontmatter}


\section{Introduction}

The development of modern electronics has been
fundamentally shaped by the semiconductor diode, which enables unidirectional
current flow through nonreciprocal resistance~\cite{r1}. Recently, a quantum
analogue known as the superconducting diode effect~\cite{r2} has emerged, where a
material exhibits zero resistance (supercurrent) in one direction but becomes
resistive in the opposite. This phenomenon offers a step toward ultra-low
dissipation circuits, with applications ranging from energy-efficient
computing to directionally selective quantum sensors~\cite{r3,r4}. The fundamental
requirement for realizing superconducting diode effect is the simultaneous
breaking of spatial inversion symmetry and time-reversal symmetry. While these
symmetries are traditionally broken using external magnetic fields and
spin-orbit coupling, recent research has focused on achieving a zero-field
superconducting diode effect through intrinsic means, such as the use of van
der Waals heterostructures~\cite{r5}, chiral superconductors~\cite{r6}, or trapped
magnetic flux~\cite{r7}. Also, the realization of the Josephson diode effect (JDE)
has emerged as a cornerstone in the pursuit of nonreciprocal superconducting
electronics. By enabling a direction-dependent critical current, the JDE
provides a dissipationless analog to the semiconductor diode, presenting a
pathway toward high-efficiency rectification and logic elements in quantum
circuits~\cite{r8}. Recent works include JDEs in supercurrent
interferometers~\cite{r9n}, asymmetric inline long Josephson
junctions~\cite{r10n}, and kinked Rashba nanochannels~\cite{r11n}, as
well as a dynamical route to supercurrent nonreciprocity based on
parametric driving~\cite{r12n}. While initial research focused on static methods to break
spatial inversion and time-reversal symmetry, such as the use of
non-centrosymmetric materials~\cite{r9}, recent attention has shifted toward
dynamical symmetry breaking via time-dependent driving. A particularly powerful
procedure for achieving this is the application of biharmonic excitations. By
driving a Josephson junction with a composite signal containing two
commensurate frequencies, typically $\omega$ and $2\omega$, the system's
temporal symmetry can be precisely manipulated. The nonreciprocity in these
systems is fundamentally governed by the relative amplitude and the relative
phase of the two driving tones. Remarkably, these types of biharmonic
excitations have also been used with the goal of achieving a single electron
excitation above the Fermi sea with a minimal creation of electron-hole
excitations in tunnel junctions (TJs)~\cite{r10}, which is of great interest to
quantum computing.

In two recent works~\cite{r11,r12}, the diode effect induced by applying a two-tone
driving with relative phase $\theta$, amplitudes $I_{1}$ and $I_{2}$, and
frequencies $\omega$ and $2\omega$, respectively, to a conventional
superconducting tunnel-junction is experimentally and numerically
investigated. Thus, Scheer \textit{et al.}~\cite{r11} considered the dynamics of the
superconducting phase $\varphi$ across a Josephson junction in the overdamped
limit described by%
\begin{align}
\frac{\hslash}{2eR}\overset{.}{\varphi}+I_{c}\sin\left(  \varphi\right)   &
=I_{dc}+I_{ac}(t),\nonumber\\
I_{ac}(t)  &  \equiv I_{1}\cos\left(  \omega t\right)  +I_{2}\cos\left(
2\omega t+\theta\right)  , \tag{1}%
\end{align}
where $R$ and $I_{c}$ are a shunt resistance and the critical current,
respectively. In both the slow- and fast-driving regime, the authors studied
the strength of the diode effect by computing the diode efficiency
$\eta=\left\vert I_{c}^{+}+I_{c}^{-}\right\vert /\left\vert I_{c}^{+}%
-I_{c}^{-}\right\vert =\left\vert I_{ac}^{+}+I_{ac}^{-}\right\vert /\left\vert
2I_{c}-I_{ac}^{+}+I_{ac}^{-}\right\vert $, where $I_{ac}^{+}\ \left(
I_{ac}^{-}\right)  $ is the maximum (minimum) value of $I_{ac}(t)$, while they
used the estimate%
\begin{equation}
\eta\approx\frac{3I_{1}^{2}I_{2}}{32I_{c}^{3}\left(  \omega/\omega_{c}\right)
^{4}}\left\vert \cos\left(  \theta\right)  \right\vert \tag{2}%
\end{equation}
in the fast-driving regime, with $\omega_{c}\equiv2eI_{c}R/\hslash$ being the
characteristic frequency of the circuit. The authors obtained numerical estimates
of the diode efficiency and the width of the supercurrent region for the
particular values $I_{2}/I_{1}=1/2$ and $\theta=0$  (cf. Ref.~\cite{r11}).
Nevertheless, the authors did not provide any theoretical explanation 
for these special values of $I_{2}/I_{1}$ and $\theta$. On the other hand,
Borgongino \textit{et al.}~\cite{r12} studied basically the same problem as Scheer
\textit{et al.}~\cite{r11} but with
\begin{equation}
I_{ac}(t)\equiv I_{1}\sin\left(  \omega t\right)  +I_{2}\sin\left(  2\omega
t+\theta\right)  \tag{3}%
\end{equation}
instead of the biharmonic signal considered in Eq. (1). Their experimental and
numerical findings indicate that both the diode's efficiency and the asymmetry
of the supercurrent region reach their maximum values when $I_{2}/I_{1}=1/2$
and $\theta=\pm\pi/2$~\cite{r12}. In particular, they found that: (i) The critical
current in ideal rectification, $\left\vert I_{s}^{\ast}\right\vert
=I_{c}-\min\left(  I_{ac}^{+},\left\vert I_{ac}^{-}\right\vert \right)  $,
presents, as a function of the relative amplitude, an extremum at $I_{2}%
/I_{1}=1/2$ for $\theta=\pm\pi/2$ (cf. Fig. 5(d) in Ref.~\cite{r12}). (ii) The diode
efficiency $\eta$ presents, as a function of the relative amplitude, a single
maximum at $I_{2}/I_{1}=1/2$ for $\theta=\pi/2$ and several different values
of the peak-to-peak amplitude $I^{pp}$ (cf. Fig. 9(f) of Supporting
Information~\cite{r12}). Also, the authors stated that \textquotedblleft The ideal
diode was obtained by adjusting $\theta$ and the amplitude of the microwave
drive, in good agreement with what was predicted in ref 35\textquotedblright%
\ and that \textquotedblleft The maximum $I_{s}^{\ast}$ is achieved for
$I_{2}/I_{1}=0.5$, as predicted by perturbation theory,$^{35}$%
\textquotedblright\ (cf. Ref.~\cite{r12}), where \textquotedblleft ref
35\textquotedblright\ is Ref.~\cite{r11}. However, the authors did not present any
theoretical argument for these special values of $I_{2}/I_{1}$ and
$\theta$, since such an argument is not given in Ref.~\cite{r11} either.

Another quantum-mechanical phenomenon occurring in TJs, noise spectroscopy of
a quantum tunnel junction with a biharmonic voltage drive, has been
experimentally and numerically studied in two recent works~\cite{r10,r13} with the
ultimate goal of obtaining phase-coherent electronics for quantum computing.
Thus, Gabelli \textit{et al.}~\cite{r10} reported experimental measures of shot
noise (\textit{variance} of the current fluctuations) in a TJ under a dc
voltage $V_{dc}$ and a biharmonic excitation%
\begin{equation}
V_{ac}(t)=V_{ac1}\cos\left(  \omega t\right)  +V_{ac2}\cos\left(  2\omega
t+\varphi\right)  , \tag{4}%
\end{equation}
with $\omega\equiv2\pi\nu$, $h\nu\gg k_{B}T,T=70$ mK, and high frequency
$\nu=10$ GHz (quantum regime). They characterized the effects of the
biharmonic excitation by calculating the noise spectral density $S_{2}$ from
the corresponding nonequilibrium electron distribution function and showed
that \textquotedblleft adding an excitation at frequency $2\nu$ [$2\omega$]
with the proper amplitude and phase can \textit{reduce} the noise of the
junction excited at frequency $\nu$ only.\textquotedblright\ (cf.
Ref.~\cite{r10}).
Commenting on their experimental and numerical results, the authors claimed
that, \textquotedblleft For experimental parameters $T=0.14h\nu/k_{B}$ and
$eV_{ac1}=5.4h\nu$, we obtain that optimal values are $eV_{ac2}=eV_{dc}%
=2.4h\nu$ and $\varphi=0$. For $\varphi=\pi$, the waveform is reversed [see
Fig. 4(a)] and the minimum occurs at the opposite value of $V_{dc}$. Figure
4(b) shows noise measured for $eV_{ac2}=2.7h\nu$ (i.e. close to optimal)... .
All the data (symbols) are very well fitted by the theory (solid
lines).\textquotedblright\ (cf. Fig. 4(b) in Ref.~\cite{r10}), and that
\textquotedblleft It is interesting to remark that the waveform we found that
minimizes the noise for a given $V_{ac1}$ at finite temperature is not close
to Lorentzian, but corresponds almost to the first two harmonics of a
Lorentzian with a dc offset [see Fig. 1(b)].\textquotedblright\ (cf. Appendix
B and Fig. 6 in Ref.~\cite{r10}). However, the authors did not supply any
theoretical reason for these special values of $V_{ac2}%
/V_{ac1}\simeq1/2$ and $\varphi=\left\{  0,\pi\right\}  $. Almost
simultaneously, Vanevic \textit{et al.}~\cite{r13} analyzed the experimental data
obtained by Gabelli \textit{et al.}~\cite{r10} to show how the excess
photon-assisted noise in the presence of an in-phase second harmonic is
composed of the contributions of electron-hole pairs created, while a suitable
choice of the biharmonic excitation waveform [Eq. (4)] optimally decreases the
probability of the electron-hole pair leading to the minimal excess noise
observed in Ref.~\cite{r10} (cf. Ref.~\cite{r13}). Specifically, Vanevic \textit{et al.}
calculated the current noise power at low temperature $T\ll\omega$:
\begin{equation}
S=GF\sum_{n=-\infty}^{\infty}\left\vert eV_{dc}+n\hbar \omega\right\vert \left\vert
a_{n}\right\vert ^{2}, \tag{5}%
\end{equation}
where $G$ and $F$ are the conductance and the Fano factor, respectively, and
the excess noise
\begin{equation}
S_{ac\mid N}=2GF\hbar\omega\sum_{n=1}^{\infty}n\left\vert a_{-N\mp n}\right\vert
^{2}, \tag{6}%
\end{equation}
where $N$ is an integer and the upper (lower) sign is taken for $N\geq0$
$\left(  N<0\right)  $, and where
\begin{equation}
a_{n}\equiv\sum_{m=-\infty}^{\infty}J_{n-2m}(eV_{ac1}/(\hbar \omega))J_{m}\left(
eV_{ac2}/(2\hbar \omega)\right)  , \tag{7}%
\end{equation}
with $J_{n}\left(  z\right)  $ being the Bessel function of the first kind.
Also, at $eV_{dc}/(\hbar \omega)=N$, $S_{ac\mid N}$ is given by the sum of the
probabilities of electron-hole pair creations $p_{k}^{(N)}$:%
\begin{equation}
S_{ac\mid N}=2GF\hbar\omega\sum_{k}p_{k}^{(N)} \tag{8}%
\end{equation}
(cf. Ref.~\cite{r13}). Commenting on their numerical results, the authors claimed
(for $\varphi=0$, $eV_{dc}/(\hbar \omega)=3$, and $eV_{ac1}/(\hbar
\omega)=5.4$) that (they took $\hbar=1$)
\textquotedblleft For the amplitude $eV_{ac2}/ \omega\approx2.6$, the total
probability $p_{1}+p_{2}$ of the electron-hole pair creation exhibits a
minimum. This leads to the minimal excess noise in Eq. (2) [Eq. (8)] which has
been observed in Ref.~\cite{r8}.\textquotedblright(cf. Fig. 4 in
Ref.~\cite{r13}), where
\textquotedblleft Ref. 8\textquotedblright\ is a preprint of
Ref.~\cite{r10}.
Nevertheless, the authors did not provide any theoretical argument for these
special values of $V_{ac1}/V_{ac2}\approx2$ and $\varphi=0$.

In this work, we discuss how the law of RU~\cite{r14,r15,r16}
 provides a well-reasoned and unified explanation of such experimental
and numerical results (cf. Refs.~\cite{r10,r11,r12,r13}). This law establishes that there
exists a universal excitation waveform that optimally enhances directed
ratchet transport (DRT) by critically breaking the generalized time-reversal
symmetry and the generalized parity symmetry and refers to the criticality
scenario that emerges when such symmetries are broken, regardless of the
nature of the dynamic equation in which the breaking of such symmetries
results in DRT. Notably, the universal excitation waveform also tends to
maximize transport coherence because it forces the available energy to be used
primarily for translation rather than spreading (the entities that are
transported become maximally `locked' into the driving excitation), and hence
the variance of the velocity distribution is expected to be minimized. 

Thus, when the universal excitation waveform drives a quantum system
(like cold atoms in an optical lattice, electrons in a TJ, or a
 Bose-Einstein condensate) to its absolute maximun average
velocity, the resulting suppression of momentum fluctuations has
fundamental implications for the Heisenberg's uncertainty principle.
Indeed, when the universal excitation waveform shrinks the momentum
uncertainty to a razor-thin minimum, the uncertainty in the position
must increase to satisfy Heisenberg's inequality. As particles achieve
perfectly synchronized, maximum-momentum directed transport with
virtually no momentum spread, their wave packets spatially spread out.
This means that the particles reach a highly delocalized, coherent
state, i.e., the universal excitation waveform drives the system
toward the Heisenberg limit, creating a highly stable quantum phase
current with the minimum possible noise allowed by the laws of nature.
Since in quantum technologies, such as superconducting qubits and
Josephson TJs, the randomness governed by Heisenberg's uncertainty
principle manifests as shot noise, the law of RU allows for the
optimization of shot noise suppression in the current (momentum space).
This exceptional noise reduction provides an ideal mechanism for the
efficient generation of nonclassical photonic states, including highly 
optimized squeezed states of light. Regarding driven-junction
phenomena, the above is the expected scenario in the simplest physical
situation: overdamped regime and absence of thermal noise. At a finite
temperature and overdamped regime, as stablished by the RU law,
thermal noise acts as an effective noise-induced alteration in the
height of the potential barrier which is in turn controlled by the DSB
mechanism~\cite{r22}. In an overdamped Josephson junction, thermal noise and
quantum noise typically compete to degrade the coherence and
efficiency of the device. However, RU provides the exact mathematical
condition required to maximize the macroscopically rectified current
(counteracting the thermal suppression of the barrier) while
simultaneously acting as an optimal filter that suppresses excess
quantum noise (minimizing the energy wasted on creating random quantum
fluctuations and electron-hole pairs). This dual optimization makes
the RU law highly relevant for quantum technologies, specifically in
superconducting integrated power electronics, electron quantum optics,
and the development of high-fidelity quantum computing components. In
the underdamped regime of a Josephson junction (finite capacitance),
however, rectification becomes highly nonlinear and non-adiabatic.
While the RU law still defines the optimal foundation for breaking the
relevant space-time symmetry~\cite{r14}, thermal noise can induce
transitions between running, locked, and chaotic states. The system's
kinetic energy acts as a rolling memory, i.e., the particle's current
state dependes upon the history of the entire wave cycle. This memory
allows for unexpected phenomena, such as chaos and current reversals,
and introduces a time-lag (phase delay) between the application of the
biharmonic force and the actual velocity of the particle, and hence
the optimal values of the relative initial phase deviate from the
universal ones associated with the overdamped regime~\cite{r14}. On the
other hand, the universal optimal value of the relative amplitude only
remains valid in the adiabatic limit, i.e., when the driving period is
the longest timescale of the system (the driving frequency is much
lower than the plasma frequency), thus allowing the Josephson junction
to behave as if it were overdamped.

The law
of RU has elucidated previous experimental results concerning DRT of atoms in
a Hamiltonian quantum ratchet (the values of the parameter used, which were
selected to maximize DRT, correspond to those of the universal biharmonic
waveform)~\cite{r17}, fluxons in uniform annular Josephson junctions in one case and
of cold atoms in dissipative optical lattices in another, both driven by
biharmonic fields~\cite{r18}, pure spin currents in organic
materials~\cite{r19},
superparamagnetic colloidal particles exposed to a magnetic-field-induced
asymmetric sawtooth potential~\cite{r20}, and topological charge pumping in a
Floquet-Bloch band using a harmonic lattice potential~\cite{r21}. It is worth noting
that this law has elucidated the historical problem of the rectification of
the Brownian particle by biharmonic excitations~\cite{r22}, has explained the
maximum strength of DRT in complex networks of driven damped
pendula~\cite{r23}, and
has been numerically confirmed in the context of skyrmion
ratchets~\cite{r24}, among
many other cases.

\section{Superconducting tunnel junction}

Since Eq. (1) describes a driven
overdamped system, the violation of the time-reversal symmetry implies the
breaking of the biharmonic excitation's symmetry $I_{ac}(-t)=-I_{ac}(t)$,
while the dc term $I_{dc}$ breaks \textit{per se} this symmetry. For
$I_{dc}=0$, the optimal values predicted from RU~\cite{r14} to obtain maximal
dc-voltage across the junction $V\equiv\hslash\left\langle \overset{.}%
{\varphi}\right\rangle /(2e)$ are%
\begin{equation}
\left(  I_{2}/I_{1}\right)  _{opt}=1/2,\theta_{opt}=\left\{  0,\pi\right\}  .
\tag{9}%
\end{equation}
For $I_{dc}>0$ $(I_{dc}<0)$, the degree-of-symmetry-breaking (DSB)
mechanism~\cite{r14} predicts that the optimal relative phase $\theta_{opt}=\pi$ $\left(
\theta_{opt}=0\right)  $ yields maximal transport directed in the opposite
direction to the dc term, while the value $\theta_{opt}=0$ $\left(
\theta_{opt}=\pi\right)  $ does in the same direction as the dc term, which
explains the asymmetric zero-voltage regions in the \textit{IV}-curves (cf.
Fig. 1(c) in Ref.~\cite{r11}). Also, the DSB mechanism predicts that the extension
of these zero-voltage regions is maximal when $I_{2}/I_{1}=\left(  I_{2}%
/I_{1}\right)  _{opt}=1/2$. This prediction is fully confirmed by the
experimental results shown in Figs. 5(c) and 5(e) in Ref.~\cite{r12} for the
biharmonic excitation given by Eq. (3). Indeed, for the choice of Eq. (3) and
$I_{dc}=0$, the optimal values predicted from RU to obtain maximal dc-voltage
across the junction are now~\cite{r14}%
\begin{equation}
\left(  I_{2}/I_{1}\right)_{opt}=1/2,\theta_{opt}=\pm\pi/2, \tag{10}%
\end{equation}
where the two signs $\pm$ correspond to directed transport in opposite
directions, which also explain both the behavior of the critical current in
ideal rectification $\left\vert I_{s}^{\ast}\right\vert $ as a function of the
relative amplitude (cf. Fig. 5(d) in Ref.~\cite{r12}) and the maximum of the diode
efficiency $\eta$ at $I_{2}/I_{1}=\left(  I_{2}/I_{1}\right)  _{opt}=1/2$ for
$\theta=\theta_{opt}=\pi/2$ and several different values of the peak-to-peak
amplitude $I^{pp}$ (cf. Fig. 9(f) in Supporting Information~\cite{r12}).
Furthermore, the optimal values $\theta_{opt}=\pm\pi/2$ [Eq. (10)] explain the
extrema of $\eta$ as a function of $\theta$ (cf. Fig. 3(f) in
Ref.~\cite{r12}).
Notice that the dependence of $\eta$ on the amplitudes in Eq. (2) (cf.
Ref.~\cite{r11}) does not agree with the prediction from RU [cf. Eqs. (9) and (10)].
Clearly, the reason comes from the assumption that the contributions of the
amplitudes $I_{1}$ and $I_{2}$ to the DRT are independent, contrary to what
the law of RU states. To further clarify this point, let us consider the
following reparameterization of the biharmonic drivings in Eqs. (1) and (3):%
\begin{align}
\frac{I_{ac}(t)}{I_{0}}  &  =f_{c,c,\alpha,\theta}\left(  t\right)
\equiv\zeta\cos\left(  \omega t\right)  +\alpha\left(  1-\zeta\right)
\cos\left(  2\omega t+\theta\right)  ,\tag{11}\\
\frac{I_{ac}(t)}{I_{0}}  &  =f_{s,s,\alpha,\theta}\left(  t\right)
\equiv\zeta\sin\left(  \omega t\right)  +\alpha\left(  1-\zeta\right)
\sin\left(  2\omega t+\theta\right)  , \tag{12}%
\end{align}
where $I_{0}$ is an amplitude, $\zeta\in\left[  0,1\right]  $ accounts for the
relative amplitude of the two harmonic components, while $\alpha>0$ is an
amplitude prefactor. According to the law of RU, the optimal value of $\zeta$
comes from the condition that the amplitude of the odd harmonic component must
be twice that of the even harmonic component in Eqs. (11) and (12):%
\begin{equation}
\zeta_{opt}=\zeta_{opt}\left(  \alpha\right)  \equiv\frac{2\alpha}{1+2\alpha},
\tag{13}%
\end{equation}
while the diode efficiency $\eta$ [cf. Eq. (2)] should now scale as%
\begin{equation}
\eta\approx C\alpha\zeta^{2}\left(  1-\zeta\right)  , \tag{14}%
\end{equation}
where $C=C\left(  I_{0},I_{c},\omega/\omega_{c},\theta\right)  $ is
independent of $\zeta$. Equation (14) indicates that $\eta$ presents a single
maximum at $\zeta_{\max}=2/3$, irrespective of the particular value of the
prefactor $\alpha$, while the coincidence of $\zeta_{\max}=2/3$ with
$\zeta_{opt}\left(  \alpha=1\right)  $ is purely accidental (cf. Appendix in
Ref.~\cite{r25}). Thus, the predictions from the RU law explain, in a general
framework, the aforementioned
special values of $I_2/I_1$ and $\theta$ found by Scheer \textit{et al.} 
(cf. Fig. 3 in Ref.~\cite{r11}) and the
experimental results obtained by Borgongino \textit{et al.} (cf. Figs. 3(g)
and 9(f) in Ref.~\cite{r12} Supporting Information). In contrast to the prediction
of Eq. (14), a numerical evaluation of the driving asymmetry 
$\eta_{ac}=\frac{\left (I_{ac}^{+}+I_{ac}^{-}\right )}{\left
(I_{ac}^{+}-I_{ac}^{-}\right )}$ with $I_{ac}(t)$ given by
Eq. (11) confirm the RU prediction [Eq. (13)] over a wide range of $\alpha$
values (see Fig. 1). Equation (1) was integrated by using a
fourth-order Runge-Kutta method with time steps chosen in the range
0.001-0.005, randomly chosen initial conditions, and integration
typically up to $10^4$ drive cycles for each fixed set of parameters
($I_c$=1.18 {$\mu$}A, $R$=2.15 $\Omega$, $\omega=0.01\omega_c$ in Figs. 1
and 3). The time-averaged voltage was tipically calculated over $10^5$
drive cycles after the transients had been removed.
Note that $\eta_{ac}=\eta\left| \frac{2
I_c}{I_{ac}^{+}-I_{ac}^{-}}-1 \right|$.
\begin{figure}[htb]
\centering
\includegraphics[width=0.6\textwidth]{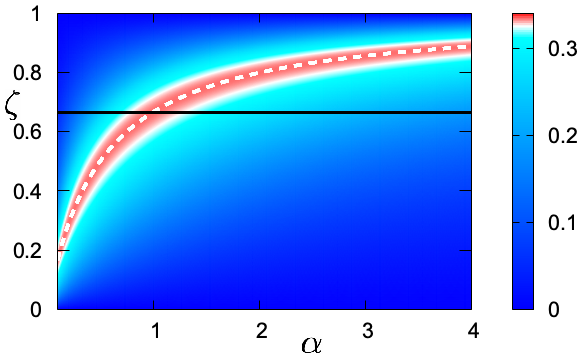}
\caption{
Driving asymmetry  
$\eta_{ac}=\frac{I_{ac}^{+}+I_{ac}^{-}}{I_{ac}^{+}-I_{ac}^{-}}\equiv 
2 D_{c,c,\alpha,\theta=0}(\zeta)$ [cf. Eqs. (A8) and (A9)]
versus prefactor
$\alpha$ and relative amplitude $\zeta$ for $\theta=0,I_{dc}
=0$ [cf. Eq. (11)]. For each ($\alpha,\,\zeta$) the amplitude $I_0$
is tuned to $I_c/I_{ac}^+$ in order to achieve $\eta=1$ (ideal diode). 
Theoretical
predictions for the maximum value from RU [cf. Eq.(13)]
and Eq.(14) are shown as dashed and solid lines, respectively.
}
\label{fig1}
\end{figure}

\begin{figure}[htb]
\centering
\includegraphics[width=0.8\textwidth]{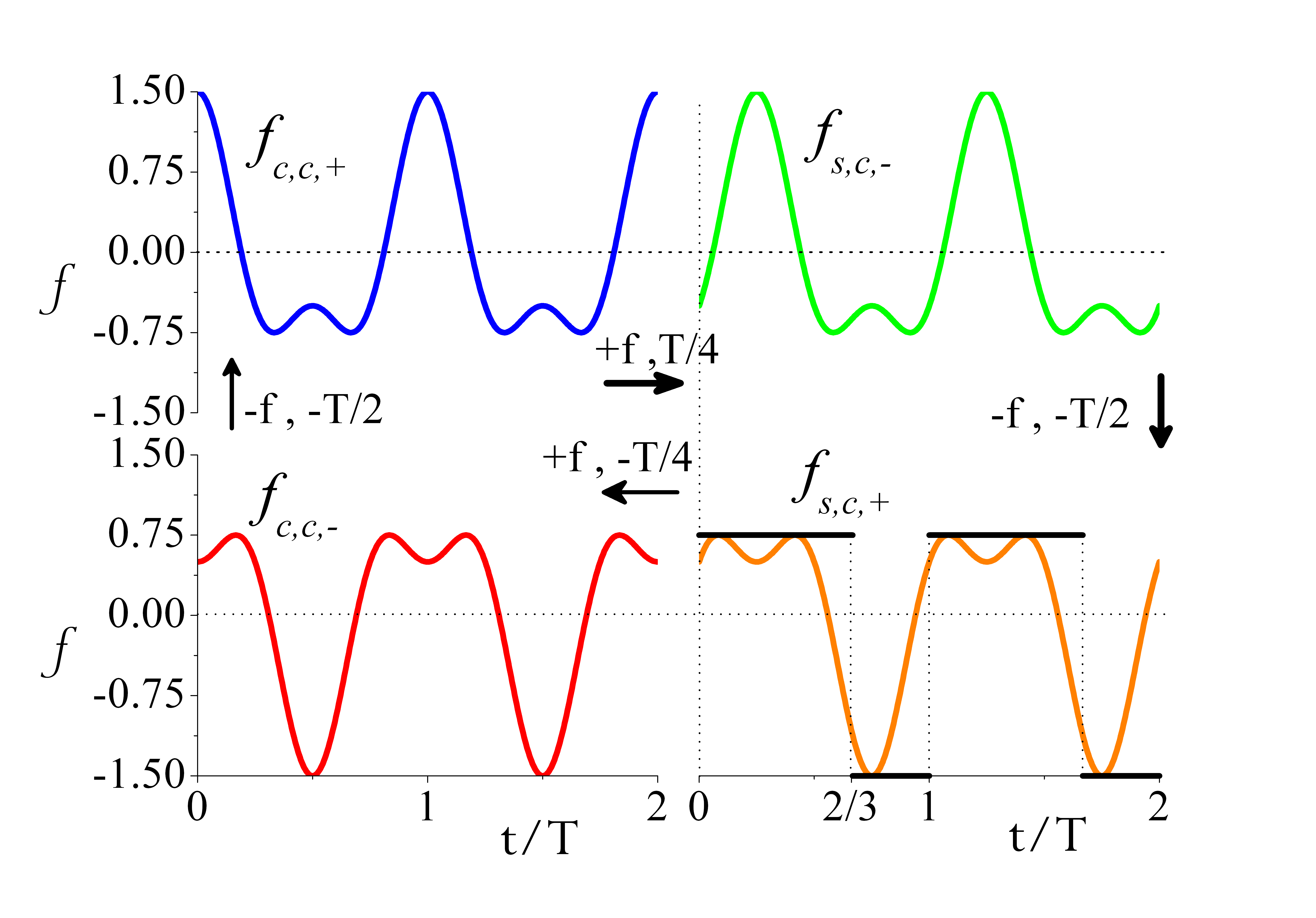}
\caption{Functions $f_{c,c,\pm}(t)\equiv\cos\left(  \omega t\right)
\pm(1/2)\cos\left(  2\omega t\right)  $ and $f_{s,c,\pm}(t)\equiv\sin\left(
\omega t\right)  \pm(1/2)\cos\left(  2\omega t\right)  $ [cf. Eqs. (9)-(12)]
representing the optimal biharmonic waveform vs $t$ over two periods. The
dichotomous waveform (thick solid line) represents the exact universal
excitation waveform, while horizontal and vertical arrows indicate the
symmetries that relate the different biharmonic representations.}
\label{fig2}
\end{figure}

Figure 2 shows the four equivalent expressions of the
(single) optimal biharmonic waveform associated with the above optimal values
[cf. Eqs. (9) and (10)], while the numerical results shown in Fig. 3 confirm
the dependence of the VI curves on the relative amplitude according to the RU
law (cf. Fig. 5(f) in Ref.~\cite{r12}). In particular, the upper boundary of the
region wherein the junction exhibits zero average voltage is well fitted by
the function representing the effective $\zeta$-dependent `load' term arising
from the renormalization of the biharmonic excitation (cf. Fig. 3; see
the Appendix).

\begin{figure}[htb]
\centering
\includegraphics[width=0.47\textwidth]{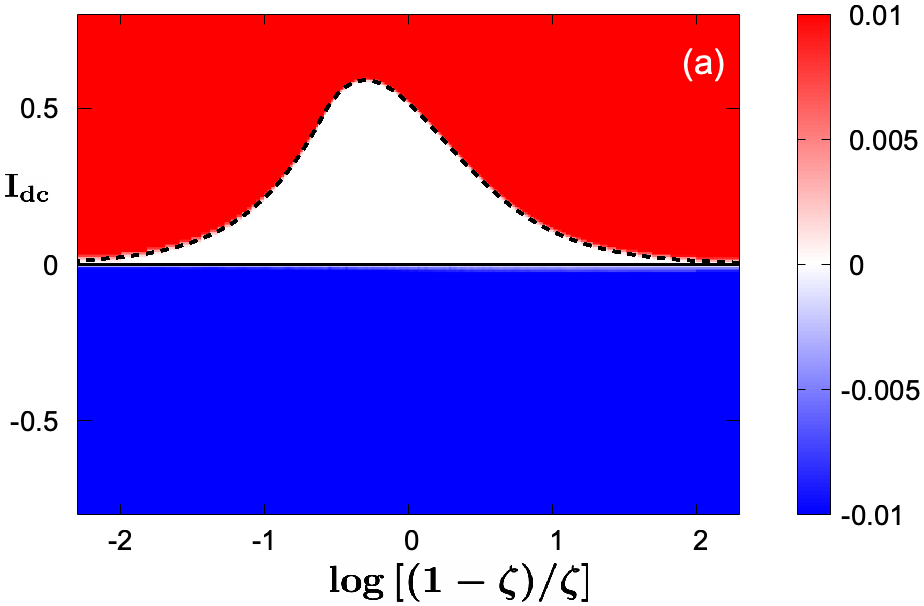}
\includegraphics[width=0.47\textwidth]{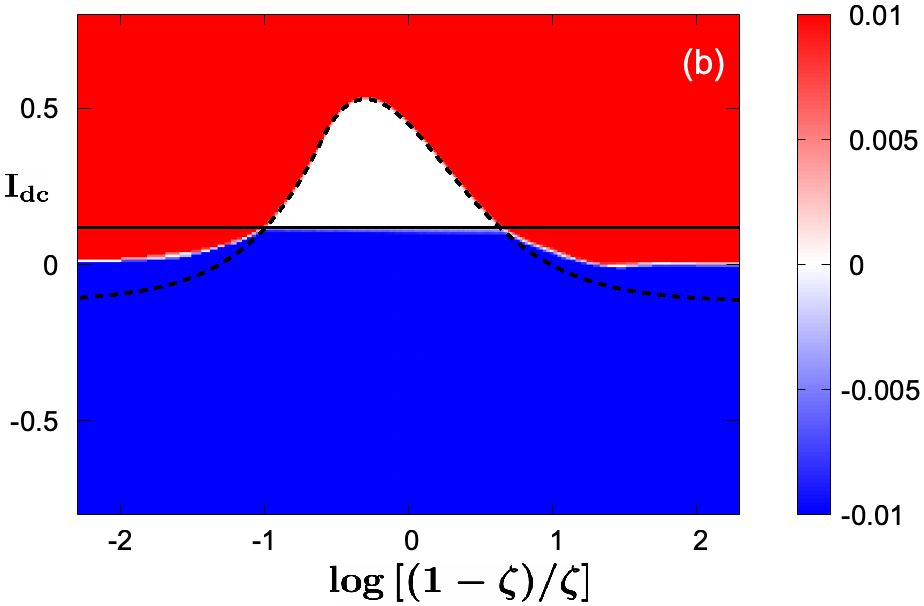}
\caption{Josephson junction voltage $V_{JJ}$ (in ${\mu}V$) versus
$I_{dc}$ (in ${\mu}A$) and $\log\left[
\left(  1-\zeta\right)  /\zeta\right]  $ [cf. Eqs. (1) and (12)] for
$\theta=\theta_{opt}=\pi/2$, $\alpha=1$, $R=2.15\,\Omega$,
$\omega=0.01\omega_c$, and $I_c=1.18\, {\mu}A$.  (a)
$I_{0}=1.18\,{\mu}A$ ($\eta=1$, ideal diode).
 (b) $I_0=1.3\,{\mu}A$ (when the ratchet effect occurs: $V_{JJ}<0$ for
$I_{dc}>0$). 
The
theoretically predicted boundaries (dashed line: upper bound, solid
line: lower bound) of the superconducting channel are also
plotted in both cases [cf. Eq. (A10) in the Appendix]. 
Colorbar is truncated to $[-0.01,0.01]$ to highlight the
zero-crossing region. Values below $-0.01\,{\mu}V$ and above $0.01\,{\mu}V$ are
saturated in uniform blue and red, respectively.}

\label{fig3}
\end{figure}

\section{Shot noise in TJs}

For the biharmonic excitation considered in Ref.~\cite{r10}
[cf. Eq. (4)], the optimal values predicted from RU to obtain a maximum
average current in the junction, and therefore a maximum reduction in the
variance of electron population fluctuations (shot noise) are%
\begin{equation}
\left(  V_{ac2}/V_{ac1}\right)  _{opt}=1/2,\varphi_{opt}=\left\{
0,\pi\right\}  , \tag{15}%
\end{equation}
where the optimal relative phase $0$ $\left(  \pi\right)  $ applies when
$V_{dc}>0$ $\left(  V_{dc}<0\right)  $ according to the DSB
mechanism~\cite{r14}.
This means that the number of electron-hole pairs for $\varphi=\varphi
_{opt}=0$ must decrease (increase) as compared to its number for $\varphi
=\pm\pi/2$ when $V_{dc}>0$ $\left(  V_{dc}<0\right)  $ because the normalized
ratcheting effect corresponding to the biharmonic excitation Eq. (4) is
equivalent to an $\zeta$-dependent `load' term that helps to produce a greater
(smaller) excess of electrons along with (due to its opposition to) the dc
voltage (see the Appendix). Indeed, these optimal values [Eq. (15)] correspond
to the optimal waveform (cf. Fig. 2) that Gabelli \textit{et
al.}~\cite{r10} found
minimizes \textquotedblleft the noise for a given $V_{ac1}$ at finite
temperature\textquotedblright\ (cf. Appendix B in Ref.~\cite{r10}), which explains
their aforementioned experimental results, in particular the greater
efficiency of the optimal biharmonic excitation compared to Lorentzian pulses.

\begin{figure}[htb]
\begin{center}
\includegraphics[width=0.4\textwidth]{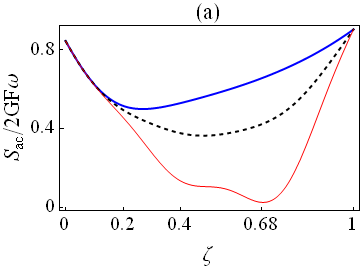}
\includegraphics[width=0.4\textwidth]{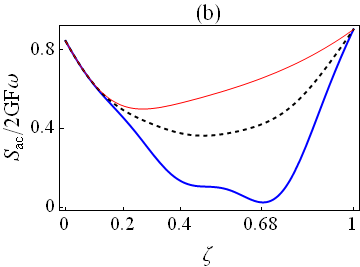}\\
\includegraphics[width=0.4\textwidth]{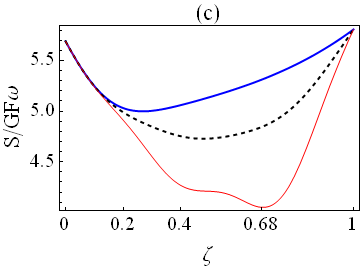}
\includegraphics[width=0.4\textwidth]{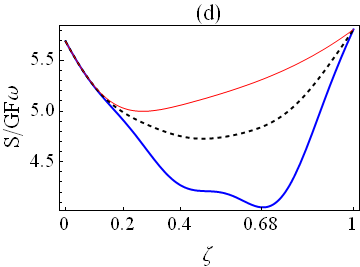}\\
\end{center}
\begin{center}
\includegraphics[width=0.4\textwidth]{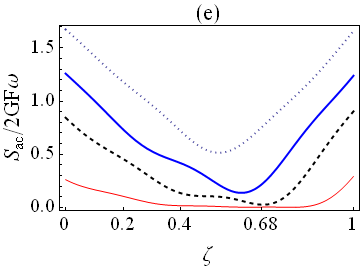}
\end{center}
\caption{(a), (b) Excess noise $S_{ac}$ [cf. Eq. (6)] and (c), (d) current
noise $S$ [cf. Eq. (5)] as a function of the relative amplitude $\zeta$ for
$eV_{ac}/(\hbar \omega)\equiv e\left(  V_{ac1}+V_{ac2}\right)  /(\hbar
\omega)=8.1$ [cf. Eq.
(4)] and three values of the relative phase: $\varphi=\varphi_{opt}\equiv0$
(thin line), $\varphi=\pi/2$ (dashed line), and $\varphi=\varphi_{opt}%
\equiv\pi$ (thick line). Versions (a) and (c) ((b) and (d)) correspond to
$eV_{dc}/(\hbar \omega)=4$ ($eV_{dc}/(\hbar \omega)=-4$). (e) Excess
noise $S_{ac}$ as a function of $\zeta$ for $\varphi=\varphi_{opt}=0$ 
and four values of $eV_{dc}/(\hbar \omega)$: 2 (dotted
line), 3 (thick line), 4 (dashed line) and 6 (thin line). We have used $\hbar$=1 in
all panels.}
\label{fig4}
\end{figure}

On the other hand, the experimental results of the excess noise $S_{ac}$ as a
function of the dc voltage for $\varphi=0$ (cf. Fig. 2(b) in
Ref.~\cite{r13})
indicate that $S_{ac}\approx0$ at $eV_{dc}/(\hbar \omega)=4$. Note that $eV_{dc}%
/(\hbar \omega)=\left\{  4,-4\right\}  $ are the values at which the difference
between biharmonic and monoharmonic photon-assisted noise $\Delta
S_{2,ac}(V_{dc})\equiv S_{2,ac}(V_{dc},V_{ac1},V_{ac2})-S_{2,ac}%
(V_{dc},V_{ac1},V_{ac2}=0)$ exhibits maxima at $\varphi=\varphi_{opt}%
\equiv\left\{  \pi,0\right\}  $, respectively, (cf. Fig. 4(c) in
Ref.~\cite{r10})
because in such cases the equivalent $\zeta$-dependent `load' term is in
opposition to the dc voltage, thus yielding a maximum relative increase of the
photon-assisted noise. Thus, after using the reparameterization $V_{ac}\equiv
V_{ac1}+V_{ac2}$, $\zeta\equiv V_{ac1}/(V_{ac1}+V_{ac2})$ in Eqs. (5)-(7) and
taking $eV_{ac}/(\hbar \omega)=8.1$ consistently with the parameters
used in Ref.~\cite{r13}, 
one obtains that both
the current noise $S$ and the excess noise $S_{ac}$ present an absolute
minimum at $\zeta=\zeta_{\min}=0.6858\simeq\zeta_{opt}\equiv2/3$ for
the corresponding optimal initial phases depending upon the sign of the dc
voltage $eV_{dc}/(\hbar \omega)=\pm4$ [Eq. (15); see Fig. 4], as
predicted from RU. The analysis of the series in Eqs. (5)-(7) with
 {\it{Mathematica}} indicates
 their rapid convergence (it was sufficient to take 40 terms for $S$ 
 and $a_n$, and 20 terms for $S_{ac|N}$).
 We found that $S_{ac}$ exhibits a global (flat) minimum at ($eV_{dc}/(\hbar
 \omega)\geq 4, \zeta=\zeta_{min}\approx\zeta_{opt}\equiv 2/3$). Thus,
 one obtains $\zeta_{min}=0.6858$ for $eV_{dc}/(\hbar \omega)=4$ and
 $\zeta_{min}=0.6333$ for $eV_{dc}/(\hbar \omega)= 6$,
 while $S_{ac}\approx 0$ at the flat minimum (see Fig.
 4(e)).
 
Therefore, Lorentzian waveforms should be replaced with the optimal biharmonic
waveform with a suitable dc offset in the quantization condition $\int_{0}%
^{T}V(t)dt=Nh/e$ to transport $N$ electrons cleanly in the most efficient way
when $eV_{dc}/(\hbar \omega)$ is sufficiently small (cf. Fig. 6 in
Ref.~\cite{r10}).

\section{Conclusions}

We have shown how the interplay between the RU law and Heisenberg's
uncertainty principle 
provides a well-reasoned and unified explanation of previous experimental and
numerical findings concerning the diode effect in conventional superconducting
tunnel-junctions in one case, and controlled suppression of electron-hole pair
generation in a tunnel junction in the other, both in the presence of
biharmonic driving fields. In particular, this law optimizes the rectification
efficiency of a superconducting tunnel junction and provides optimal dynamic
control of elementary excitations in quantum conductors. It is therefore
expected that this law will be especially useful in the contexts of quantum
computing (particularly for superconducting qubits like the fluxonium) and
electron quantum optics (to inject single electrons into a circuit on demand
with minimal shot noise generation). We expect the criticality scenario giving rise to RU can be discussed
similarly in the case of Josephson junctions with non-sinusoidal
current-phase relations~\cite{r26n,r27n}, since the space-time symmetries
that must be broken are the same as in the purely sinusoidal model. Indeed, applying the RU driving field optimally reduces excess
 quantum noise relative to baseline DC noise levels, thus preserving
 the fragile quantum states of the qubits, directly boosting coherence
 times and gate fidelities. Also, the RU law acts in tandem with Berry
 phase in driven lattice systems (like Floquet-Bloch bands) to
 perfectly orchestrate charge pumping~\cite{r21}. It provides a clear-cut
 prescription for optimizing quantum tunneling through
 potential barriers, allowing for fast, high-fidelity state transfer
 across the qubit processor. In fluxonium qubits, the RU driving
 field: i) Minimizes the activation barrier selectively in one
 direction. This can maximize the nonadiabatic Landau-Zener transitions
 needed for ultrafast gates while drastically suppressing leaks to
 accidental states~\cite{r28n}. ii) Optimally suppresses excess quantum shot
 noise. It can allow engineers to reset the qubit at high speeds without
 generating unwanted electron-hole pairs or quasiparticles that
 degrade future coherence~\cite{r29n}. iii) Maximizes the efficiency of the
 superconducting diode effect, which yields a massive rectification
 range for a fraction of the power. This can eaase the engineering
 constraints on individual lines, making large-scale integrated
 fluxonium arrays far more scalable.

\section*{Acknowledgment}
The authors thanks Rub\'{e}n Seoane for
interchanges about this issue. P.J.M. acknowledges financial support from the Ministerio de
Ciencia, Innovaci\'{o}n y Universidades (MICIU, Spain) through Project No.
PID2023-147734NB-I00 cofinanced by FEDER funds and
from Departamento de Industria e Innovaci\'{o}n del Gobierno de Arag\'{o}n
(FENOL group, Grant No. E36\_23R). R.C. acknowledges financial support from the
Junta de Extremadura (JEx, Spain) through Project No. GR24101 cofinanced by
FEDER funds. 

\section*{Appendix: Optimal biharmonic waveform}

According to the law of RU,
the four equivalent expressions of the (non-normalized) optimal biharmonic
excitation in the overdamped regime~\cite{r14,r15} are given by%

\begin{align}
f_{c,c,\pm}(t)  &  \equiv\cos\left(  \omega t\right)  \pm(1/2)\cos\left(
2\omega t\right)  ,\nonumber\\
f_{s,c,\pm}(t)  &  \equiv\sin\left(  \omega t\right)  \pm(1/2)\cos\left(
2\omega t\right)  , \tag{A1}%
\end{align}
which satisfy the symmetries%
\begin{align}
f_{c,c,+}(t+T/2)  &  =-f_{c,c,-}(t),\nonumber\\
f_{s,c,+}(t+T/2)  &  =-f_{s,c,-}(t),\nonumber\\
f_{c,c,\pm}(t+T/4)  &  =f_{s,c,\mp}(t). \tag{A2}%
\end{align}
Note that the effectiveness of this optimal biharmonic excitation comes from
its waveform is that of the best biharmonic (two terms) approximation to the
\textit{exact} universal excitation waveform in the sense of its Fourier
series~\cite{r16} (see Fig. 2). To obtain the normalized ratcheting effect
corresponding to the biharmonic drivings on the right-hand side of Eqs. (11)
and (12) (with the corresponding optimal values of their relative phases and a
fixed value of $\alpha$), one needs affine transformations to renormalize them
such that their waveforms change while their amplitudes and images remain
constant as the relative amplitude $\zeta$ varies from $0$ to
$1$~\cite{r15}. Thus,
for the case $f_{c,c,\alpha=1,\theta=0}\left(  t\right)  $ [cf. Eq. (11)] for
example, the normalized function reads%
\begin{align}
f_{c,c,\alpha=1,\theta=0}^{\ \ast}\left(  t\right)   &  =\frac{f_{c,c,\alpha
=1,\theta=0}\left(  t\right) \frac{2}{3} }{M_{c,c}\left(\zeta\right)-m_{c,c}\left(  \zeta\right)  }-D_{c,c,\alpha
=1,\theta=0}\left(  \zeta\right)  ,\tag{A3}\\
D_{c,c,\alpha=1,\theta=0}\left(  \zeta\right)   &
\equiv\frac{M_{c,c}\left(\zeta\right)+m_{c,c}\left(
\zeta\right)  }{2\left[  M_{c,c}\left(\zeta\right)-m_{c,c}\left(  \zeta\right)  \right]  },\tag{A4}
\end{align}
where $M_{c,c}\left(\zeta\right)$ and $m_{c,c}\left(\zeta\right)$ stand for the
maximum and minimum, repectively, of $f_{c,c,\alpha=1,\theta=0}$:
\begin{align}
M_{c,c}\left(\zeta\right) & \equiv 1,\;\; \forall \zeta \tag{A5}\\
m_{c,c}\left(  \zeta\right)   &  \equiv %
\begin{cases}
\frac{9\zeta^{2}-16\zeta+8}{8\zeta-8},& \;\zeta\leq4/5\\
&\\
1-2\zeta,&\;4/5\leq\zeta
\end{cases} \tag{A6}%
\end{align}
One finds that the corresponding maximal transmitted impulse over a
half-period,%
\begin{align}
I\left[  f_{c,c,\alpha=1,\theta=0}^{\ \ast}\left(  t\right)  \right]  \left(
\zeta\right)   &  \equiv\left\vert \int_{T/2}f_{c,c,\alpha=1,\theta=0}%
^{\ \ast}\left(  t\right)  dt\right\vert \tag{A7}\\
&  =\pi D_{c,c,\alpha=1,\theta=0}\left(  \zeta\right)  ,\nonumber
\end{align}
presents a single maximum at $\zeta=2/3$, as predicted by the RU
law~\cite{r15}.
Similarly, one straightforwardly obtains the normalized functions for the
other three biharmonic expressions, which include the \textit{common} (except
for sign, as expected) $\zeta$-dependent `load' (constant excitation) term:%
\begin{align}
D_{c,c,\alpha,\theta=0}\left(  \zeta\right)   &  =-D_{c,c,\alpha,\theta=\pi
}\left(  \zeta\right)  ,\nonumber\\
D_{s,s,\alpha,\theta=\pi/2}\left(  \zeta\right)   &  =-D_{s,s,\alpha
,\theta=-\pi/2}\left(  \zeta\right) \nonumber\\
&  =D_{c,c,\alpha,\theta=\pi}\left(  \zeta\right)  , \tag{A8}%
\end{align}
with%
\begin{equation}
D_{s,s,\alpha,\theta=\pi/2}\left(  \zeta\right)  \equiv%
\begin{cases}
\frac{\zeta\left (8 \alpha\left (\zeta-1 \right )+\zeta \right
)}{2\left (-4\alpha\left(\zeta-1\right )+\zeta\right )^2},&\;\zeta\leq\frac{4\alpha
}{1+4\alpha}\\
&\\
\frac{\alpha (\zeta-1)}{2\zeta},&\;\zeta \geq\frac{4\alpha}{1+4\alpha}.\\
\end{cases}  \tag{A9}\\%
\end{equation}
Therefore, the maximal impulse transmitted by the four normalized biharmonic
excitations over a half-period is the same: that given by Eq. (A7) [cf. Eqs.
(A7) and (A8)], as expected from the criticality scenario giving rise to
maximally enhanced DRT~\cite{r14,r15,r16}.

In the context of the superconducting diode in the adiabatic limit,  
the theoretical boundaries for the superconducting
channel ($V_{JJ}=0$) [see Figs (3a) and (3b)] when a junction with
critical current $I_c$  is driven by a dc-current $I_{dc}$ and an ac-current
 $I_{ac}(t)$ [cf. Eq. (12)] are:
\begin{equation}
\begin{cases}
\text{lower bound:}\,\,\,\,\,\,&-I_c - I_0\,m_{s,s}\left(\zeta\right)\\
&\\
\text{upper bound:}\,\,\, & I_c+I_0\frac{M_{s,s}\left(\zeta\right)+m_{s,s}\left(\zeta\right)}{2
D_{s,s,\alpha=1,\theta=\pi/2}}-I_0\,m_{s,s}\left(\zeta\right)
\end{cases}\tag{A10}
\end{equation}
where we must take into account that
\begin{align}
M_{s,s}&=-m_{c,c},\nonumber\\
m_{s,s}&=-M_{c,c}, \tag{A11}
\end{align}
for $f_{s,s,\alpha,\theta=\pi/2}$ [cf. Eq. (12)] and
$f_{c,c,\alpha,\theta=0}$ [cf. Eq. (11)].


\begin{thebibliography}{99}                               
%

\bibitem {r1}W. Shockley, Bell Syst. Tech. J. \textbf{28}, 435 (1949).

\bibitem {r2}J. Hu, C. Wu, and X. Dai, Phys. Rev. Lett. \textbf{99}, 067004 (2007).

\bibitem {r3}M. Nadeem, M. S. Fuhrer, and X. Wang, Nat. Rev. Phys. \textbf{5},
558 (2023).

\bibitem {r4}J. Ma, R. Zhan, and X. Lin, Adv. Physics Res. \textbf{4}, 2400180 (2025).

\bibitem {r5}J. Xiong, J. Xie, B. Cheng, Y. Dai, X. Cui, L. Wang, Z. Liu, J.
Zhou, N. Wang, X. Xu, X. Chen, S.-W. Cheong, S.-J. Liang, and F. Miao, Nat.
Commun. \textbf{15}, 4953 (2024).

\bibitem {r6}T. Le, Z. Pan, Z. Xu, J. Liu, J. Wang, Z. Lou, X. Yang, Z. Wang,
Y. Yao, C. Wu, and X. Lin, Nature \textbf{630}, 64 (2024).

\bibitem {r7}N. Jiang, M. Maeda, Y. Yamaguchi, M. Watanabe, M. Tokuda,
K. Takaki,S. Masaki, T. Ikushima, and T. Koyanagi, Phys. Rev. B \textbf{112}, 235313 (2025).

\bibitem {r8}J. Clarke and F. K. Wilhelm, Nature \textbf{453}, 1031 (2008).

\bibitem{r9n} R. Seoane Souto, M. Leijnse, and C. Schrade, Phys. Rev.
Lett. \textbf{129}, 267702 (2022).

\bibitem{r10n} C. Guarcello, S. Pagano, and G. Filatrella, Appl. Phys.
Lett. \textbf{124}, 162601 (2024).

\bibitem{r11n} A. Maiellaro, M. Trama, J. Settino, C. Guarcello, F.
Romeo, and R. Citro, SciPost Phys. \textbf{17}, 101 (2024).

\bibitem{r12n} Y. Yerin, S.-L. Drechsler, A. A. Varlamov, and F.
Giazotto, Phys. Rev. Lett. \textbf{137}, 056002 (2026).

\bibitem {r9}F. Ando, Y. Miyasaka, T. Li, J. Ishizuka, T. Arakawa, Y. Shiota,
T. Moriyama, Y. Yanase, and T. Ono, Nature \textbf{584}, 373 (2020).

\bibitem {r10}J. Gabelli and B. Reulet, Phys. Rev. B \textbf{87}, 075403 (2013).

\bibitem {r11}D. Scheer, R. Seoane Souto, F. Hassler, and J. Danon, New J.
Phys. \textbf{27}, 033013 (2025).

\bibitem {r12}L. Borgongino, R. Seoane Souto, A. Paghi, G. Senesi, K.
Skibinska, L. Sorba, E. Riccardi, F. Giazotto, and E. Strambini, Nano Lett.
\textbf{25}, 14451 (2025).

\bibitem {r13}M. Vanevic and W. Belzig, Phys. Rev. E \textbf{86}, 241306(R) (2012).

\bibitem {r14}R. Chac\'{o}n, J. Phys. A \textbf{40}, F413 (2007).

\bibitem {r15}R. Chac\'{o}n, J. Phys. A \textbf{43}, 322001 (2010);
\textbf{54}, 209501 (2021).

\bibitem {r16}R. Chac\'{o}n and P. J. Mart\'{\i}nez, Int. J. Bifurcation Chaos
\textbf{31}, 2150109 (2021).

\bibitem {r22}P. J. Mart\'{\i}nez and R. Chac\'{o}n, Phys. Rev E \textbf{87},
062114 (2013); \textbf{88}, 019902 (2013); \textbf{88}, 066102 (2013).

\bibitem {r17}T. Salger, S. Kling, T. Hecking, C. Geckeler, L. Morales-Molina,
and M. Weitz, Science \textbf{326}, 1241 (2009).

\bibitem {r18}R. Chac\'{o}n and P. J. Mart\'{\i}nez, Phys. Rev. E \textbf{104},
014120 (2021).

\bibitem {r19}H. Li, T. Gao, and S. Xie, J. Phys. Chem. Lett. \textbf{16}, 8906 (2025).

\bibitem {r20}J. Mart\'{\i}n-Roca, L. Izquierdo Solis, F. Mart\'{\i}nez
Pedrero, P. Casadejust, I. Pagonabarraga, and C. Calero, Phys. Rev. Lett.
\textbf{135}, 028301 (2025).

\bibitem {r21}R. Chac\'{o}n and P. J. Mart\'{\i}nez, Phys. Lett. A
\textbf{577}, 131461 (2026).


\bibitem {r23}R. Chac\'{o}n, A. Mart\'{\i}nez Garc\'{\i}a-Hoz, P. J.
Mart\'{\i}nez, and D. Dur\'{a}n, Phys. Rev E \textbf{111}, 034205 (2025).

\bibitem {r24}W. Chen, L. Liu, Y. Ji, and Y. Zheng, Phys. Rev. B \textbf{99},
064431 (2019).

\bibitem {r25}P. J. Mart\'{\i}nez and R. Chac\'{o}n, Nonlinear Dyn.
\textbf{111}, 12973 (2023).

\bibitem {r26n}M. J. A. Stoutimore, A. N. Rossolenko, V. V. Bolginov, V.
A. Oboznov, A. Y. Rusanov, D. S. Baranov, N. Pugach, S. M. Frokov, V.
V. Ryazanov, and D. J. Van Harlingen, Phys. Rev. Lett.
\textbf{121}, 177702 (2018).

\bibitem {r27n}B. Raes, N. Tubsrinuan, R. Sreedhar, D. S. Guala, R.
Panghotra, H. Dausy, Cl\'{e}cio C. de Souza Silva, and J. Van de Vondel, Phys. Rev. B
\textbf{102}, 054507 (2020).

\bibitem{r28n} H. Zhang, S. Chakram, T. Roy, N. Earnest, Y. Lu, Z.
Huang, D. K. Weiss, J. Koch, and D. I. Schuster, Phys. Rev. X
\textbf{11}, 011010 (2021).

\bibitem {r29n} J. Lauwens, K. Moors, and B. Sor\'{e}e, Phys. Rev. B,
\textbf{113}, 054526 (2026).

\end{thebibliography}
\end{document}